\documentclass[aps,floats,showpacs,superscriptaddress,floatfix,preprint]{revtex4}
\usepackage{epsfig}
\begin{document}
\draft

\noindent
{\bf Comment on: ``Roughness of Interfacial Crack Fronts: Stress-Weighted
Percolation in the Damage Zone''}

A recent Letter \cite{Sch_03}, by Schmittbuhl, Hansen and Batrouni (SHB) 
addresses the question of how interfacial cracks roughen in the presence
of disorder. SHB explain this process by a stress induced gradient percolation
model that takes into account the damage accumulated, and translates
that into a self-affine crack front profile. In this comment, we 
point out that the results  presented in Ref.~\cite{Sch_03} do
not prove self-affinity but rather support self-similarity of the crack
fronts. This result however would be in disagreement with  
experiments \cite{maloy}.

In the model of SHB the strain gradient induces a damage profile 
and a crack front results. As the load is raised 
the width of the front $W$ increases approximately as a power law,
and eventually saturates. As in gradient percolation \cite{SAP}, 
the saturated width $W^*$ scales with the gradient of the 
damage profile $1/l_y$ as $W^* \sim l_y^\alpha$ with $\alpha=\nu/(1+\nu)$
where $\nu$ is the correlation exponent of the underlying
percolation problem. Since in Ref.~\cite{Sch_03} $l_y \sim L_x$, where 
$L_x$ is the lattice size parallel to the front, SHB combine the 
initial dynamic scaling with the that of the saturated width into 
a ``Family-Vicsek''-like scaling form $W(L_x,t)= L_x^\alpha f(t/L_x^z)$, 
and conclude that the fronts are self-affine interfaces.
Such an attempt is misleading, since presenting data in such a
form does not imply that the fronts are self-affine. In gradient percolation
$\alpha$ can {\it not} be interpreted as a roughness exponent \cite{SAP}:
the front is self-similar (i.e. the scaling is isotropic) up to a 
lengthscale $\xi \sim W$ \cite{GAB}  and it is trivially flat 
on scales beyond $\xi$. Self-affinity implies instead that on {\it any} 
lengthscale $l<\xi$ the system rescales anisotropically.
Although strain induced correlations could change 
the values of the critical exponents from the standard percolation ones,
the basic picture remains the same.

Fig. 1 shows the data of the corresponding Figure 1 from \cite{Sch_03}, 
displaying the broken springs. We also include the hull of the (damage) 
gradient percolation cluster and the corresponding Solid-On-Solid (SOS) 
interface. Comparing these two shows that the SOS presentation is just 
an artificial projection from the fractal perimeter of the damage zone 
which it is not self-affine. In particular, we see that the size of 
overhangs is of the same order of the width. We have also studied an effective 
medium model in the spirit of Ref.~\cite{zhr} in which the strain 
profile is computed similarly to Ref.~\cite{Sch_03}, but the damage
is replaced by its average along the transverse 
direction \cite{own}. This model is able to reproduce the features of the
``Family-Vicsek'' data collapse of SHB, but the fronts
are obviously described by standard gradient percolation. 
From our simulations we find that the gradient $l_y$ depends 
on the elastic constants of the problem. 
In Ref.~\cite{Sch_03} the Green function $G_{ij}$
is normalized so that $\sum_{ij} G_{ij}/(L_x L_y)$ is constant.
Since $L_y$ is kept constant this amounts to rescaling the elastic
constant by $L_x$, producing an effective dependence of $l_y$ on $L_x$.

In conclusion, a correct interpretation in the framework of gradient
percolation of the data presented in Ref.~\cite{Sch_03} implies that fronts
are self-similar rather than self-affine. Thus the model of Ref.~\cite{Sch_03}
does not explain the roughness of planar cracks observed experimentally
\cite{maloy}.

M.J.\ Alava$^{1,2}$, S.\ Zapperi$^{2}$ 

{\small $^{1}$Laboratory of Physics,
Helsinki University of Technology, FIN-02015 HUT, 
Finland.}

{$^2$SMC-INFM, Dipartimento di Fisica,
Universit\`a ``La Sapienza'', P.le A. Moro 2
00185 Roma, Italy}

\begin{figure}[ht]
\centerline{\psfig{file=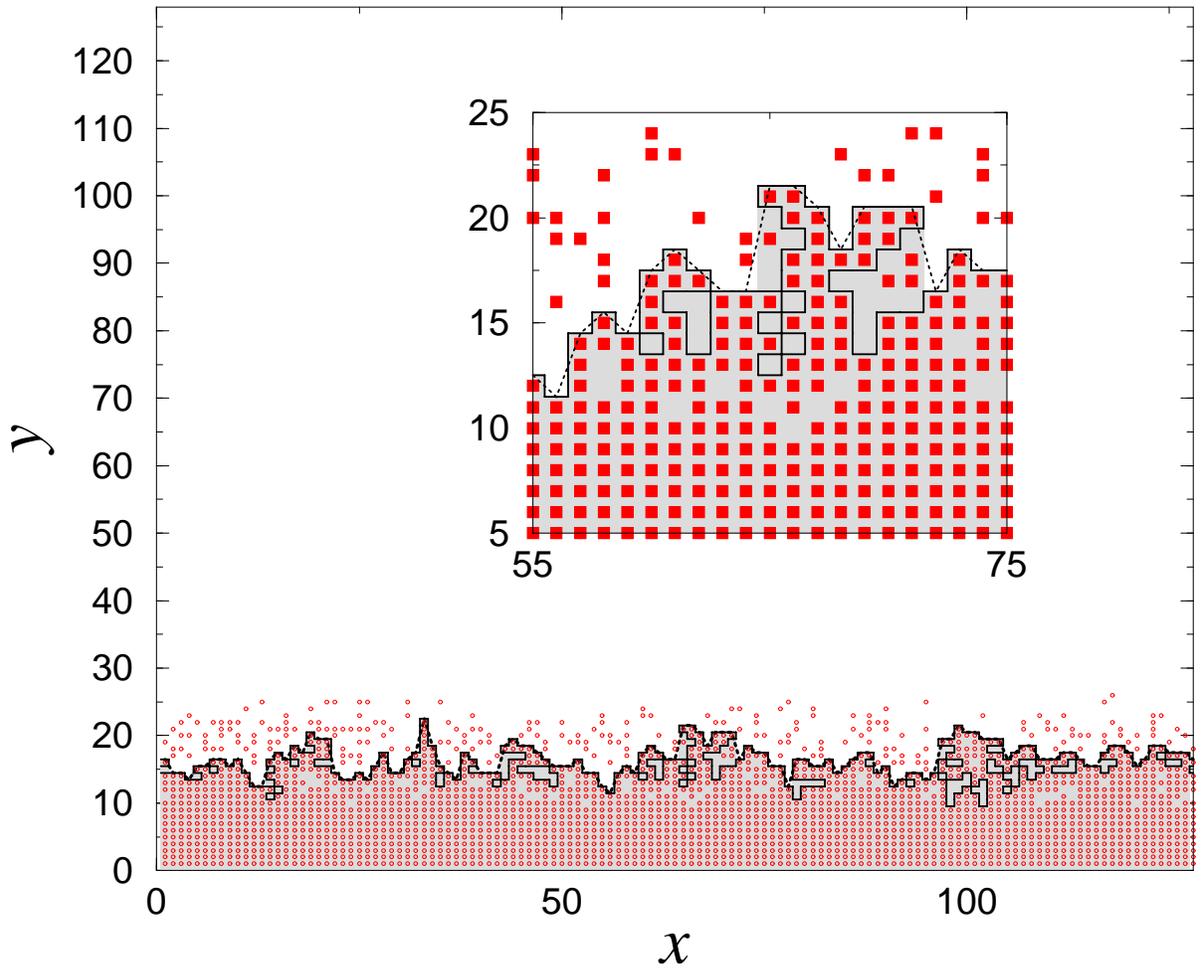,width=16cm,clip=!}}
\caption{The damage reported in Fig.~1 of Ref.~\protect\cite{Sch_03} is 
plotted together with the front perimeter (solid line) and the SOS
approximation (dotted line). As shown in the inset,
the perimeter displays substantial overhangs, whose size
is comparable with the width, and it is thus not self-affine.}
\end{figure}

\end{document}